\UseRawInputEncoding
\documentclass[%
 aps,
 prd,
twocolumn,
nofootinbib,
 amsmath,amssymb,
 aps,
]{revtex4-2}
\usepackage{natbib}
\usepackage{graphicx}
\usepackage{adjustbox}
\usepackage{dcolumn}
\usepackage{bm}
\usepackage[dvipsnames]{xcolor}
\usepackage{color, colortbl}
\usepackage{relsize}
\usepackage[mathlines]{lineno}

\usepackage{stackengine}
\stackMath
\usepackage{xcolor}
\usepackage{dcolumn}
\usepackage{bm}
\usepackage{physics}
\usepackage{comment}
\usepackage{hyperref}
\usepackage{float}
\usepackage{soul}

\begin{document}

\title{Primordial black hole-star binaries via dynamical friction}


\author{Nicolas Esser}
\email[]{nicolas.esser@ulb.be}
\affiliation{Service de Physique Th\'{e}orique, Universit\'{e} Libre
de Bruxelles}
\affiliation{Brussels laboratory of the Universe -- BLU-ULB \\CP225 Boulevard du Triomphe, B-1050 Bruxelles,
Belgium}

\date{\today}

\begin{abstract}

We study a new channel for binary system formation involving stars and stellar-mass primordial black holes (PBHs) embedded in dark matter (DM) minihalos. In this scenario, binaries form when a star passes through the DM minihalo surrounding a PBH and loses sufficient energy due to dynamical friction. The continued energy loss induced by this friction is expected to drive the resulting systems to merge rapidly. We estimate their merger rate and explore the implications for two observables: rapidly decaying X-ray binaries in the Milky Way and gravitational waves sourced by compact object mergers.
We find that, for a PBH abundance $\Omega_{\text{PBH}}/\Omega_{\text{DM}}=0.01$, this mechanism naturally produces a population of $\mathcal{O}(1)$ currently observable short-lived X-ray binaries. It also leads to a non-negligible gravitational wave event rate of $\mathcal{O}(0.3)$ $\text{Gpc}^{-3}\text{yr}^{-1}$, potentially involving high mass ratios and black holes in the lower or upper mass gap. Notably, most mergers arise in low-mass galaxies, making the latter rate sensitive to the low-mass end of the galaxy stellar mass function. The dynamical friction channel thus offers a plausible explanation for several unusual observations reported in recent years across both the X-ray and gravitational wave domains.

\end{abstract}


\maketitle

\section{Introduction}

The nature of dark matter (DM) is still unknown; while it may be composed of particles, such as WIMPs or axions (see e.g. \cite{Arcadi_2018,Chadha-Day_2021} for reviews), primordial black holes (PBHs) \cite{Zeldovich,Hawking} are another appealing DM candidate. However, extensive studies of the cosmological and astrophysical implications of the existence of primordial black holes of various masses have imposed many constraints on their abundance. They have therefore been excluded from making the entirety of the dark matter in most mass ranges \cite{Carr_review,Green_2020}. This does not imply that PBHs cannot exist; they may still constitute a subdominant fraction of the DM and could interact with the rest of it.

Scenarios in which PBHs make up a secondary component of the DM, with the majority being particlelike DM, have thus been explored by multiple authors recently (see e.g. \cite{Mack_2007,Ricotti_2007,Ricotti_2008,Ricotti_2009,Lacki_2010,Eroshenko_2016,Boucenna_2018,Bertone_2019,Eroshenko_2020,Cai_2021,Boudaud_2021,Agius_2024}). It was shown that PBHs may accrete some of the surrounding particle dark matter in the early Universe, such that part of the DM would be clustered in minihalos -- or spikes -- around PBHs. These PBHs surrounded by DM minihalos have sometimes been dubbed ``dressed'' PBHs (dPBHs) in recent studies \cite{Kavanagh_2018,Hertzberg_2020}.

The recent detection of gravitational waves (GWs) by the LIGO-Virgo collaboration \cite{LV_first,LV_1,LV_2,LV_3} has opened a new window for studying primordial black holes. Since then, extensive efforts have been made to characterize the GW signals that could arise from PBH mergers (see e.g. \cite{Raidal_2024} for a review). The possibility of ``astro-primordial'' hybrid mergers, where a PBH collides with a compact stellar remnant, has also been considered \cite{Vattis_2020,Kritos_2021,Tsai_2021,Nitz_2021,Sasaki_2022}. However, hybrid binaries must form through a dynamical channel, and their merger rate was found to be in general negligible, except in some very specific cases within clustered environments \cite{Kritos_2021}.

In the past decades, some X-ray binary events exhibiting anomalously fast orbital decay were observed in the Milky Way (MW) \cite{Gonzalez_2014}. It has been recently suggested that they could be explained if the black hole was in fact of primordial origin, and surrounded by a dark matter spike \cite{Chan_2023,Ireland_2024}. However, the possible formation mechanisms of such systems were not studied in detail.

In this work, we investigate a new formation channel for hybrid binaries, based on the presence of DM spikes surrounding PBHs. We evaluate whether this channel could be responsible for the observed fast-decaying X-ray binaries or contribute to an observable rate of GW events. More specifically, we rely on the mechanism of dynamical friction: as a star\footnote{Throughout this paper, we use the term ``star'' to refer broadly to both main-sequence stars and compact stellar remnants.} passes through the minihalo surrounding a stellar-mass PBH, it will lose energy due to the friction exerted by the DM, and may eventually form a bound binary with the PBH at the center of the DM spike. Continued energy loss from friction during subsequent minihalo crossings will cause the orbit to decay rapidly, leading to a fast merger. We compute the merger rates of such systems and apply our results to the two aforementioned observables.

We find that hybrid binaries emitting X-rays form in the Milky Way at a rate of $\Gamma_\text{XRB} \sim \mathcal{O}(10^{-7}) \text{ yr}^{-1}$. Interestingly, the typical X-ray emission lifetime of the observed systems with anomalously rapid orbital decay is approximately $1/\Gamma_\text{XRB}$ \cite{Gonzalez_2014}. The dynamical friction channel therefore offers a plausible explanation for the population of rapidly decaying X-ray binaries observed in the Milky Way.

We also obtain non-negligible volumetric merger rates of primordial black hole-stellar black hole and primordial black hole-neutron star binaries, at the level of $\mathcal{O}(0.3) \text{ Gpc}^{-3}\text{yr}^{-1}$. We therefore conclude that hybrid mergers between compact stellar objects and dressed PBHs could be responsible for some of the exotic GW events -- i.e. those with high mass ratios and/or black holes in the mass gaps -- observed in the past decade \cite{Abbott_2020,Abbott_2020_bis}. 

Note however that these rates scale linearly with the relative abundance of PBHs compared to the total DM, defined as $f_\text{PBH}\equiv\Omega_{\text{PBH}}/\Omega_{\text{DM}}$. In this work, we fix $f_\text{PBH}=0.01$, in broad agreement with current conservative constraints on stellar-mass PBHs \cite{Agius_2024}.

The rest of this paper is organized as follows. In Sec.~\ref{sec:dPBH}, we review the properties of DM minihalos surrounding PBHs and discuss their survival within galaxies. Additional details on the survival of DM spikes are provided in Appendix \ref{app:survival}. In Sec.~\ref{sec:binaryformation}, we compute the formation rate of hybrid binaries via the dynamical friction channel, as a function of the energy lost by stars during crossings of DM minihalos. We analytically and numerically evaluate this energy loss in Sec.~\ref{sec:energyloss}. In Sec.~\ref{sec:mergerrate}, we determine the fraction of newly formed hybrid binaries that will merge in a finite time and derive the corresponding merger rates. In Sec.~\ref{sec:applications}, we apply our results to two observables: fast-decaying X-ray binaries in the Milky Way and GW events. We conclude in Sec.~\ref{sec:conclusion}.

\section{Dressed primordial black holes}
\label{sec:dPBH}
\subsection{DM minihalos around PBHs}

\label{subsec:DMspikes}

Dark matter particles that fall in the deep gravitational potential of a PBH during the radiation era of the Universe may decouple from the Hubble flow and form a compact DM minihalo around the black hole \cite{Mack_2007,Ricotti_2007}. At the end of this era, the DM density in the spike follows a power-law profile $\rho_\text{sp}(r)\propto r^{-9/4}$, where $r$ is the distance to the central PBH \cite{Bertschinger_1985,Adamek_2019,Carr_2021,Jangra_2023,Ireland_2024}.

In the subsequent matter-dominated Universe, new DM shells are accreted by the minihalo, which will consequently grow in size while keeping the same density profile. Simulations \cite{Mack_2007,Ricotti_2007}, showed that this secondary infall leads to a total DM mass in the spike $m_\text{sp}\sim \mathcal{O}(100)\times m_\text{PBH}$ by redshift $z\sim30$. Due to the significant inhomogeneities of the Universe at later times, the further growth of the DM spike is uncertain. We will therefore set its mass to $m_\text{sp}=50\times m_\text{PBH}$
, which is valid up to a relative abundance $f_\text{PBH}\simeq0.01$, at which point an important fraction of the DM in the Universe must be part of minihalos around PBHs. Following \cite{Ricotti_2007,Hertzberg_2020}, the spike radius is given by
\begin{equation}
    \label{eq:radspike} r_\text{sp}\simeq1.17\left(\frac{m_\text{PBH}}{M_\odot}\right)^{1/3}\text{pc},
\end{equation}
corresponding to parsec-sized halos for the PBH masses considered in this work, i.e. around $\mathcal{O}(1-100)M_\odot$.

With the power-law density profile stated above, these values of $m_\text{sp}$ and $r_\text{sp}$ imply
\begin{equation}
\label{eq:rhospike}
    \rho_\text{sp}(r)\simeq1.8\times10^{-22}\left(\frac{r}{\text{pc}}\right)^{-9/4}\left(\frac{m_\text{PBH}}{M_\odot}\right)^{3/4}\text{g/cm}^3
\end{equation}
with a sharp cutoff at $r=r_\text{sp}$, in agreement with \cite{Hertzberg_2020} and within a factor of two of \cite{Berezinsky_2013}.
The gravitational potential of the spike $\Phi_\text{sp}(r)$ can be straightforwardly derived from this expression using the Poisson equation, and the motion of particles inside the minihalo will be dictated by the total potential $\Phi(r)=\Phi_\text{PBH}(r)+\Phi_\text{sp}(r)$, which can be written
\begin{equation}
    \label{eq:pot}
    \Phi(r)=-\frac{Gm_\text{PBH}}{r}-3Gm_\text{sp}\left(\frac{4}{3}\frac{1}{r_\text{sp}^{3/4}r^{1/4}}-\frac{1}{r_\text{sp}}\right)
\end{equation}
for $r\le r_\text{sp}$. 


\subsection{Survival of the minihalos}
While the DM spikes are formed by redshift $z\sim30$,  we will focus on their interactions with stars at recent times. It is therefore crucial to assess whether they can survive within galactic environments. Following \cite{Hertzberg_2020,Binney+Tremaine}, we consider three possible mechanisms for minihalo disruption: global tides from the galactic halo, high-velocity stellar encounters, and galactic disk shocking. Additional details are provided in Appendix \ref{app:survival}.

Due to their very high densities, dark matter minihalos are largely resilient to all three disruption mechanisms. We find that these effects are only significant in the central regions of galaxies. However, as we will show later in this paper -- and further elaborate in the Appendix -- these regions contribute only a fraction of the total merger rate induced by the dynamical friction channel. Therefore, we neglect disruption effects in the remainder of this work, as they are expected to lead to at most $\mathcal{O}(1)$ corrections.

\section{Binary formation rate}
\label{sec:binaryformation}

Since PBHs and stars do not originate from the same epoch, the only way for them to form hybrid binary systems is through dynamical channels. Typical mechanisms include three-body interactions and gravitational wave-induced energy loss. However, these mechanisms were found to be quite inefficient \cite{Kritos_2021}. DM spikes surrounding PBHs introduce a new avenue for energy loss through dynamical friction, offering a potentially more efficient formation channel for such binaries. Assuming that every PBH is surrounded by a DM minihalo, the volumetric rate of star-minihalo encounters with impact parameter between $b$ and $b+db$ and asymptotic relative velocity between $v_\infty$ and $v_\infty+dv_\infty$ is
\begin{align}
\begin{split}
    \frac{d\Gamma}{dV}&\left([b,b+db],[v_\infty,v_\infty+dv_\infty]\right)\\
    &= n_\ast n_\text{PBH} \cdot 2\pi b\cdot db \cdot v_\infty \cdot f(v_\infty) \cdot dv_\infty,
\end{split}
\end{align}
where $n_\ast$ and $n_\text{PBH}$ are the number densities of stars and PBHs, and it is assumed that the distribution $f(v_\infty)$ of relative velocities between stars and PBHs follows an isotropic Maxwell-Boltzmann law with velocity dispersion $\sigma_\text{rel}=\sqrt{2}\sigma$, where $\sigma$ is the local stellar velocity dispersion in the galactic frame.

Assuming that the asymptotic velocities are small compared to the local velocity dispersion, $v_\infty\ll \sigma_\text{rel}$, the Maxwell-Boltzmann distribution becomes uniform and one finds, with the change of variables $L=bv_\infty$ and $E=v_\infty^2/2$,

\begin{align}
\begin{split}
    \frac{d\Gamma}{dV}&\left([E,E+dE],[L^2,L^2+dL^2]\right)\\
    &=3\sqrt{6\pi}\frac{n_\ast n_\text{PBH}}{\sigma_\text{rel}^3} dL^2dE,
\end{split}
\end{align}
where $E$ and $L$ are the specific energy and angular momentum of the star-PBH system prior to collision. This result is in agreement with \cite{Press+Spergel}, up to a factor of two that was omitted in their calculations \cite{Gould_1978,Kouvaris_2007}. While the condition $v_\infty\ll\sigma_\text{rel}$ is not in general true, the collisions of interest are those leading to binary formation. As we will show in Sec.~\ref{sec:energyloss}, such collisions require low initial energies, and consequently low asymptotic velocities, thereby justifying this assumption.

During a collision, a star will lose energy due to the dynamical friction exerted by the DM particles of the minihalo. The star will end up gravitationally bound to the dressed PBH if the specific energy loss $E_\text{loss}$ (defined as a positive value here) induced by this friction exceeds its initial energy $E$,  resulting in a negative total energy when it leaves the minihalo. The total rate of binary formation in some spatial volume $V$ (e.g., a galaxy or a group of galaxies) is thus obtained as

\begin{equation}
\label{eq:BFrate}
    \Gamma_\text{BF}=3\sqrt{6\pi}\int_V\frac{n_\ast(\bm{R}) n_\text{PBH}(\bm{R})}{\sigma_\text{rel}^3(\bm{R})}dV\cdot\mathcal{I}_\text{BF},
\end{equation}
where $\bm{R}$ denotes the position vector of the binary system within the volume $V$ and
\begin{equation}
\label{eq:integral}
\mathcal{I}_\text{BF}=\iint_{E,L^2\left.\right|_{E_\text{loss}>E} } dL^2 dE    
\end{equation}
is the phase space area that leads to binary formation. The integral is performed over all energies and squared angular momenta for which the energy loss is larger than the initial energy $E$.
Importantly, the integrals over the volume and over the phase space are independent, allowing them to be treated separately in the following sections. The next section will be dedicated to the computation of $E_\text{loss}$ and of the phase space area $\mathcal{I}_\text{BF}$, while the integral over the spatial volume will be performed in Sec.~\ref{sec:applications}.

\section{Dynamical friction and energy loss}
\label{sec:energyloss}
\subsection{Dynamical friction}

The mechanism responsible for the energy loss is the dynamical friction: when a star crosses a minihalo, the slow DM particles from the halo are accelerated, while the star, in turn, is slowed down. We assume that the DM particles within the spike follow a Maxwell-Boltzmann distribution in velocity with dispersion $\bar v(r)=\sqrt{-\Phi(r)}$ -- with $\Phi(r)$ the gravitational potential from Eq.~\eqref{eq:pot} -- truncated at the escape velocity $v_\text{esc}(r)=\sqrt{2}\bar v(r)$ (in analogy with the standard halo model, see e.g. \cite{Choi_2014}). The deceleration of the star due to dynamical friction is thus \cite{Binney+Tremaine}
\begin{equation}
\label{eq:dynafric}
    \bm{a}_\text{DF}=-4\pi G^2m_\ast\rho_\text{sp}(r)\xi(r,v)\ln\Lambda\frac{\bm{v}}{v^3},
\end{equation}
where $m_\ast$ is the star mass, $\ln\Lambda=\ln\left(\sqrt{m_\text{sp}/m_\ast}\right)$ is the Coulomb logarithm \cite{Kavanagh_2020}, $\bm{v}$ is the velocity vector of the star relative to the PBH and

\begin{equation}
        \xi(r,v)=\begin{cases}
        1 & \text{if } v\geq v_\text{esc}(r)\\
        K(r,v)/K(r,v_\text{esc}(r)) & \text{if } v<v_\text{esc}(r)
    \end{cases}
\end{equation}
is the fraction of particles that are slower than the star, with
\begin{align}
\begin{split}
    K(r,v)=&-6v\exp\left(-\frac{3}{2}\frac{v^2}{\bar v(r)^2}\right)\\
    &+\sqrt{6\pi}\bar v(r)\text{Erf}\left(\sqrt{\frac{3}{2}}\frac{v}{\bar v(r)}\right).
\end{split}
\end{align}

For velocities well above or below the local minihalo velocity dispersion $\bar v(r)$, the standard results $a_\text{DF}\propto 1/v^2$ and $a_\text{DF}\propto v$ are respectively recovered \cite{Binney+Tremaine}.


\subsection{Analytical estimate of $\mathcal{I}_\text{BF}$}
\label{sec:analytical_phasespace}

The first criterion for binary formation is that the star actually crosses the minihalo. This implies the condition $L^2<L^2_\text{max}$ on the specific angular momentum, where
\begin{align}
\begin{split}
\label{eq:Lmax}
    L_\text{max}^2&=2r_\text{sp}^2\left(E-\Phi(r_\text{sp})\right)\\
    &\simeq2r_\text{sp}^2\frac{Gm_\text{dPBH}}{r_\text{sp}}
\end{split}
\end{align}
where we used the fact that the initial energy of the stars that will form binaries is negligible in front of the gravitational potential at the spike surface. We also defined the total mass of the dressed PBH as $m_\text{dPBH}=m_\text{PBH}+m_\text{sp}$.

If the star crosses the minihalo, then the second criteria for binary formation is that it loses enough energy such that its total energy becomes negative, i.e. $E<E_\text{loss}$. Assuming the star crosses the minihalo linearly with a velocity equal to the escape velocity from the surface of the spike $v\simeq\sqrt{2Gm_\text{dPBH}/r_\text{sp}}$, one can compute the specific energy loss by integrating Eq.~\eqref{eq:dynafric} along this trajectory. By further integrating the energy loss over the various possible impact parameters of the star, the mean value can then be obtained as \cite{Capela_2013,Tinyakov_2024}
\begin{align}
\label{eq:meanEloss}
    \langle E_\text{loss}\rangle&\sim 4\pi G^2m_\ast\ln\Lambda\frac{r_\text{sp}}{2Gm_\text{dPBH}}\frac{m_\text{sp}}{\pi r_\text{sp}^2}.
\end{align}

With the expressions \eqref{eq:Lmax} and \eqref{eq:meanEloss}, the integrals over energy and angular momentum in Eq.~\eqref{eq:integral} are now independent and $\mathcal{I}_\text{BF}$ is directly obtained as
\begin{equation}
\label{eq:analytical_phasespace}
    \mathcal{I}_\text{BF}\sim\langle E_\text{loss}\rangle\times L_{\text{max}}^2=4G^2m_\text{sp}m_\ast\ln\Lambda.
\end{equation}

\subsection{Numerical computation of $\mathcal{I}_\text{BF}$}
\label{subsec:numericalI}

To compute $\mathcal{I}_\text{BF}$ with more accuracy, we numerically solve the equation of motion of a test particle (representing the star) crossing the minihalo: $\bm{\dot v}=-\bm{\nabla}\Phi+\bm{a}_\text{DF}$, using Eqs.~\eqref{eq:pot} and \eqref{eq:dynafric}. In principle, the initial conditions are set by two parameters, namely the orbital energy and angular momentum $(E,L)$. Equivalently, these can be expressed in terms of the Keplerian periastron and apastron $(r_\text{min},r_\text{max})$. A key observation is that, when $r_\text{min}\ll r_\text{max}$, the energy loss is primarily controlled by $r_\text{min}$, which determines how deeply the star penetrates into the halo. For simplicity, we therefore focus on zero-energy orbits ($E=0$), corresponding to $v_\infty=0$ at $r_\text{max}=\infty$. We then solve the equation of motion within the DM spike for various initial conditions, given by the remaining parameter $r_\text{min}$, to numerically map $E_\text{loss}(r_\text{min})$. 

\begin{figure}[t]
\includegraphics[width=1.\columnwidth]{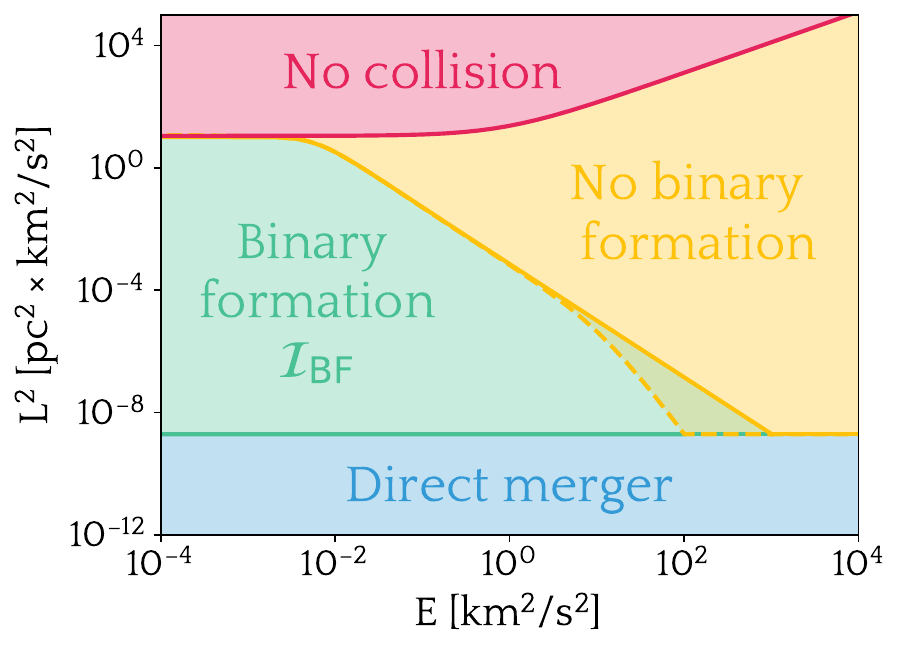}
\caption{\label{fig:cap_region} Regions of the phase space $(E,L^2)$ corresponding to different types of star-dPBH encounters, for a $10M_\odot$ PBH and a $1M_\odot$ star. In the upper (red) region, the star does not enter the minihalo. In the bottom (blue) region, there is a direct collision between the star (of solar radius $r_\odot$) and the central PBH. In the right (yellow) region, the star crosses the minihalo but does not lose sufficient energy to become gravitationally bound. In the left (green) region, the star passes through the minihalo and loses enough energy to form a bound binary system with the PBH. The dashed line indicates the boundary between regions where binary formation does and does not occur, computed without assuming that the energy loss depends only on $r_\text{min}$.}
\end{figure}

Using standard relations between orbital parameters (see, e.g., \cite{LandauLifshitz1976Mechanics}, or any other classical mechanics textbook), $r_\text{min}$ can be expressed as a function of $E$ and $L$, allowing the energy loss to be evaluated for any combination of orbital parameters. The region of phase space that allows for binary formation corresponds to all combinations satisfying  $E<E_\text{loss}(E,L)$. An example case with $m_\text{PBH}=10M_\odot$ and $m_\ast=1M_\odot$, is shown in Fig.~\ref{fig:cap_region}.

In Fig.~\ref{fig:cap_region}, the upper (red) region corresponds to trajectories for which the star does not cross the minihalo, i.e. for which $r_\text{min}>r_\text{sp}$ (which is the same condition as Eq.~\eqref{eq:Lmax}). The lower (blue) region describes trajectories for which $r_\text{min}<r_\odot$, a negligible fraction of the phase space where the star and the PBH directly merge. The central regions describe the minihalo-crossing trajectories, which lead (left, green) or do not lead (right, yellow) to binary formation.

We also solved the equation of motion with initial conditions given by the pair $(r_\text{min},r_\text{max})$ (or equivalently $(E,L)$) without using the approximation that $E_\text{loss}$ depends only on $r_\text{min}$. We found that a negligible fraction of the phase space that led to binary formation in the approximate case would not lead to binary formation in the exact case, as depicted by the yellow dashed line in Fig.~\ref{fig:cap_region}. We neglect this small fraction in the remainder of this work.

Additionally, we can now verify that the energies of binary-forming systems are indeed very small, thus justifying the earlier assumption of small asymptotic velocities, $v_\infty\ll\sigma_\text{rel}$ (see Sec.~\ref{sec:binaryformation}). As observed in the figure, the majority of the binary-forming phase space has $E=v_\infty^2/2\ll 1\text{km}^2/\text{s}^2$, meaning that as long as $\sigma_\text{rel}\gtrsim\mathcal{O}(\text{km}/\text{s})$ the approximation holds.

\begin{figure}[t]
\includegraphics[width=1.\columnwidth]{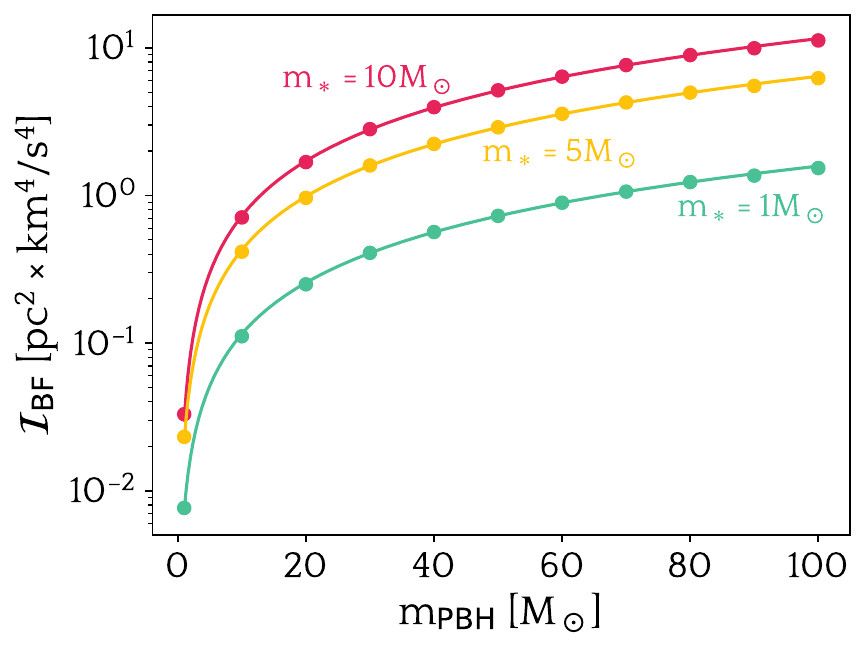}
\caption{\label{fig:cap_various} Area $\mathcal{I}_\text{BF}$ of the binary-forming phase space region, as a function of the PBH mass, for three stellar masses: $1M_\odot$, $5M_\odot$ and $10M_\odot$. The numerical results are shown as dots, while the analytical estimates from Eq.~\eqref{eq:analytical_phasespace} are displayed as continuous lines.}
\end{figure}

The integral $\mathcal{I}_\text{BF}$ is simply obtained as the area of the binary-forming region in Fig.~\ref{fig:cap_region}. We numerically calculated this area for various combinations of PBH and stellar masses. The results are displayed as dots in Fig.~\ref{fig:cap_various}.

We also used the analytic estimate from Eq.~\eqref{eq:analytical_phasespace} to compute $\mathcal{I}_\text{BF}$ for the different PBH and stellar masses, which are shown as lines in Fig.~\ref{fig:cap_various}. Interestingly, the estimate proves to be very accurate. This can be explained by the fact that the trajectories contributing most significantly to the phase space area, near the upper limit of the binary formation region, are those with large angular momenta, for which the approximation of linear trajectories crossing the minihalo is accurate.

\section{Merger rate}
\label{sec:mergerrate}

Using the results of Sec.~\ref{sec:binaryformation} and ~\ref{sec:energyloss}, one can compute the binary formation rate (Eq.~\eqref{eq:BFrate}) induced by the dynamical friction channel. However, not all formed binaries will necessarily merge. For a merger to occur, the star must continue to lose energy until its orbit decays sufficiently for it to collide with the PBH. Once the star is in a bound orbit, further energy loss is driven by dynamical friction during successive crossings of the DM spike. However, among the binaries that do form, some may have orbits with very large apastrons. In such cases, two limiting factors may prevent a merger: (\textit{i}) the time between successive crossings may be too long, preventing the star from losing enough energy to merge with the PBH within a finite time, and (\textit{ii}) the star may encounter nearby stellar objects -- which we refer to as ``perturbers'' -- that can disrupt the orbit and unbind the star from its minihalo-crossing trajectory, thereby halting the energy-loss process and ultimately preventing a merger.

Using the numerical results from Sec.~\ref{subsec:numericalI}, we obtained, for each initial binary-forming $(E,L)$ combination, the pair $(r_\text{min},r_\text{max})$ of periastrons and apastrons \textit{after} the initial collision with the minihalo. In the following subsections, we will quantify the ``sinking time'' condition and incorporate the effect of perturbers. We will then further restrict the phase space integral $\mathcal{I}_\text{BF}$ to include only those trajectories whose $(r_\text{min},r_\text{max})$ lead to a merger, in order to compute the phase space of mergers $\mathcal{I}_\text{merger}$ and the merger rate $\Gamma_\text{merger}$.

\subsection{Sinking time}
The first requirement for binaries formed via the dynamical friction channel to eventually merge is that they lose energy sufficiently rapidly. Otherwise, the star and the PBH may not collide within a finite time. Between subsequent crossings, the star follows Keplerian orbits around the minihalo. With each crossing, the star loses energy, causing the apastron to decrease while the periastron remains roughly constant, at least for the first few orbits that dominate the sinking time. The evolution of the apastron is linked to the energy evolution, driven by $E_\text{loss}(r_\text{min})$, through the equation $E=-Gm_\text{dPBH}/(r_\text{min}+r_\text{max})$. Combining this with Kepler's third law, one finds the time taken by the star to sink and merge with the PBH as a function of its initial apastron. Imposing that the sinking time is smaller than the age of the Universe $T_u\simeq10^{10}$ yr, one can invert this relation to find an upper limit on the apastron that follows the initial collision \cite{Esser_2022,Tinyakov_2024},

\begin{equation}
\label{eq:rcrittime}
    r_\text{max}<r_\text{crit}(E_\text{loss})=\frac{E_\text{loss}^2T_u^2}{2\pi^2Gm_\text{dPBH}}.
\end{equation}
where $E_\text{loss}$ is the specific energy loss, as a function of $r_\text{min}$, computed in Sec.~\ref{subsec:numericalI}. 

By plugging the mean energy loss from Eq.~\eqref{eq:meanEloss} into Eq.~\eqref{eq:rcrittime}, one can estimate the critical apastron. For a $10M_\odot$ PBH and a $1M_\odot$ star, this yields $r_\text{crit} \simeq 260 \text{ pc}$.

\subsection{Perturbers}
\label{sec:perturbers}

As we just learned, binaries that merge within a finite time can have orbit sizes of several tens to even hundreds of parsecs. During such large orbits, numerous perturbers may pass near the minihalo or the orbiting star, interacting with the system and potentially altering its orbit or even disrupting the binary. If these interactions cause the star to move into an orbit that no longer crosses the DM spike, the energy loss process will stop, preventing the system from merging.

Given the very low orbital velocity of the star -- even at the surface of the minihalo, where $v_\text{esc}(r_\text{sp})\lesssim 1\text{km}/\text{s}$, which is much smaller than the typical stellar velocity dispersions in galaxies -- we can use the impulse approximation. In this approximation, the effect of a passing perturber is considered instantaneous, meaning it only affects the velocities of the binary members, not their positions. We further assume that the perturber follows a linear trajectory with constant velocity equal to its asymptotic velocity, and passes at a distance $b$ from one of the binary members, say the star. The distance between the perturber and the (static) star at any time $t$ is then described by $r(t) = \sqrt{b^2 + (\sigma_\text{rel} t)^2}$, where $t=0$ corresponds to the point of closest approach of the perturber. For simplicity we have assumed here that all perturbers have the same velocity, equal to the relative velocity dispersion $\sigma_\text{rel}$. A more detailed treatment would involve perturbers with asymptotic velocities following a Maxwell-Boltzmann distribution with a velocity dispersion of $\sigma_\text{rel}$, which could lead to corrections of $\mathcal{O}(1)$, but these are neglected in this work. We also fix the perturber mass to the mean stellar mass, $m_p=\bar{m}_\ast\equiv0.4M_\odot$. The velocity gained by the star is then of order
\begin{equation}
    \Delta v\sim\int_{-\infty}^{+\infty}\frac{Gm_p}{r(t)^2}dt=\frac{\pi G m_p}{b\sigma_\text{rel}},
\end{equation}
while the changes in specific energy and angular momentum (with respect to the PBH) are given by $\Delta E\sim\Delta v^2/2$ and $\Delta L\sim r_\text{max}\Delta v$, respectively.

The number of perturbers passing within an impact parameter $b$ during the star's orbit of period $T$ is given by $N(b)=\pi b^2 n_\ast\sigma_\text{rel}T$. Requiring $N<1$ fixes the critical impact parameter $b_\text{min}$, below which there is less than one encounter per orbit. On the other hand, perturbers with typical interaction time $t_\text{int}\sim b/\sigma_\text{rel} > T$ will act adiabatically on the binary system and therefore will not transfer any energy. This gives a maximum value of the impact parameter $b_\text{max}$ above which there is no energy input. The total energy imparted by perturbers to the star during one orbit is then obtained by summing the contributions of all perturbers as follows:
\begin{align}
\begin{split}
    \Delta E_\text{tot}&\sim\int_{b_\text{min}}^{b_\text{max}}\frac{dN}{db}\Delta Edb\\
    &=\frac{\pi^3n_\ast T G^2 m_p^2}{\sigma_\text{rel}}\ln\left(\frac{b_\text{max}}{b_\text{min}}\right)
\end{split}
\end{align}
where $T$ is related to $r_\text{max}$ through Kepler's third law.

The total angular momentum transferred by perturbers is not computed in the same way as the total energy, since individual angular momentum contributions can be either positive or negative. For a population of binaries, the net angular momentum imparted by perturbers averages to zero. However, this population will acquire a nonzero dispersion in angular momentum, characterized by
\begin{align}
\begin{split}
\sigma_L&\sim\left[\int_{b_\text{min}}^{b_\text{max}}\frac{dN}{db}\Delta L^2db\right]^{1/2}\\
    &=Gm_pr_\text{max}\left[\frac{2\pi^3n_\ast T}{\sigma_\text{rel}}\ln\left(\frac{b_\text{max}}{b_\text{min}}\right)\right]^{1/2}.
\end{split}
\end{align}
For simplicity, we adopt this dispersion as the effective total angular momentum imparted to a binary system, i.e. $\Delta L_\text{tot}=\sigma_L$. Note the weak logarithmic dependence of the critical impact parameters on the injected energy and angular momentum. This implies that the results are not significantly altered by using more precise values for $b_\text{min}$ and $b_\text{max}$.

A star leaving the minihalo with orbital parameters $(E,L)$ will be subject to perturbers and return with updated values $(E+\Delta E_\text{tot},L+\Delta L_\text{tot})$. Consequently, its periastron will increase by an amount $\Delta r_\text{min}$, which can be computed using the standard relations between $r_\text{min}$, $E$ and $L$. This shift means the star will penetrate less deeply into the minihalo during subsequent passages, resulting in reduced energy loss. This effect must be taken into account in the energy loss calculation. Additionally, the energy input from external impulses must also be included when evaluating the net energy loss between successive crossings. The energy loss function computed in Sec.~\ref{sec:energyloss} can be modified to account for these effects in the following way:

\begin{equation}
\label{eq:newEloss}
    E_\text{loss}^\text{new}(r_\text{min})=E_\text{loss}(r_\text{min}+\Delta r_\text{min})-\Delta E_\text{tot}
\end{equation}
where we again defined the energy lost by the system as a positive value, and conversely, the energy gained as negative. With this modification, cases where perturbers remove stars from minihalo-crossing orbits are automatically accounted for, as they will yield $E_\text{loss}=0$. Perturbers thus act as a ``brake'' on the sinking of the star toward the PBH. It is important to note that this slowing effect is stronger in environments with high stellar densities and low velocity dispersions.


\subsection{Phase space of mergers}
\label{sec:mergerphasespace}

Using the results from the two previous subsections, we can now define the phase space leading to mergers as

\begin{equation}
\label{eq:integral_mergers} \mathcal{I_\text{merger}}(\bm{R})=\iint_{E,L^2\left.\right|_\text{merger}}
\end{equation}
where we now integrate over all $E$ and $L^2$ such that $E_\text{loss}>E$ and $r_\text{max}<r_\text{crit}(E_\text{loss}^\text{new}) $. In other words, we further restrict the binary formation phase space $\mathcal{I}_\text{BF}$ (Eq.~\eqref{eq:integral}) to include only orbits whose apastron lies below the critical radius given by Eq.~\eqref{eq:rcrittime}, updated to include the braking effects from external perturbers as described in Eq.~\eqref{eq:newEloss}. Since this updated energy loss depends on the local environment -- through $n_\ast(\bm{R})$ and $\sigma_\text{rel}(\bm{R})$, both of which vary with galactic position $\bm{R}$ -- it is no longer possible to separate the phase space integral from the spatial volume integral (Eq.~\eqref{eq:BFrate}). Therefore, the total merger rate is obtained as

\begin{equation}
\label{eq:mergerrate}
    \Gamma_\text{merger}=3\sqrt{6\pi}\int_V\frac{n_\ast(\bm{R}) n_\text{PBH}(\bm{R})}{\sigma_\text{rel}^3(\bm{R})}\mathcal{I}_\text{merger}(\bm{R})dV.
\end{equation}

\begin{figure}[t]
\includegraphics[width=1.\columnwidth]{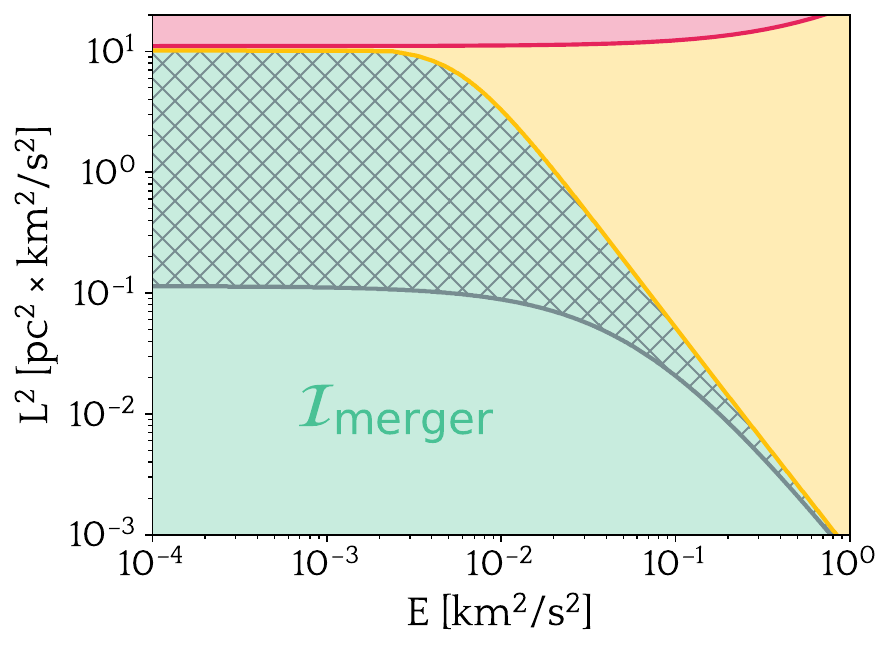}
\caption{\label{fig:cap_region_zoomin} Regions of the initial phase space $(E,L^2)$ for a $10M_\odot$ PBH and $1M_\odot$ star. The colored regions are as described in Fig.~\ref{fig:cap_region}. The shaded region corresponds to binaries that do not merge in a finite time, taking into account the braking effect of perturbers, assuming typical values of $n_\ast=1\text{ pc}^{-3}$ and $\sigma_\text{rel}=200\text{ km}/\text{s}$. The remaining portion of the green region is the phase space that will ultimately lead to mergers between stars and PBHs.}
\end{figure}

In Fig.~\ref{fig:cap_region_zoomin}, we illustrate the different types of encounters as a function of the initial orbital parameters $E$ and $L^2$, for a $10M_\odot$ PBH and a $1M_\odot$ star. As in Fig.~\ref{fig:cap_region}, the upper (red) region corresponds to trajectories where the star does not cross the minihalo. The other regions represent minihalo-crossing trajectories: those that do not lead to binary formation (right, yellow), and those that do (left, green). The shaded region marks the part of phase space that leads to binary formation, but where the binaries fail to merge within a finite time, with the braking effect of perturbers taken into account assuming $n_\ast=1 \text{ pc}^{-3}$ and $\sigma_\text{rel} = 200 \text{ km}/\text{s}$. The remaining portion of the green region represents the phase space leading to mergers, whose area is $\mathcal{I}_\text{merger}$. Note the logarithmic scale of the plot; only a subdominant fraction of formed binaries ultimately end up merging.

\section{Applications: X-ray and gravitational wave events}
\label{sec:applications}

We now apply the results of the previous sections to two observables: X-ray binaries in the Milky Way and gravitational wave events sourced by compact object mergers. To obtain merger rates in these two cases, one must choose the volume $V$ -- along with associated $n_\ast(\mathbf{R})$, $n_\text{PBH}(\mathbf{R})$ and $\sigma(\mathbf{R})$ -- over which Eq.~\eqref{eq:mergerrate} is integrated. These quantities strongly depend on galaxy modeling, as well as stellar modeling and distribution, which remain highly uncertain and subject to significant uncertainties. Therefore, the merger rates obtained in this section should be interpreted as order-of-magnitude estimates.

\subsection{X-ray binaries in the Milky Way}
\label{sec:Xray}

X-ray binaries are systems in which X-ray emission arises from the accretion of stellar material from a donor star onto a compact object. These systems have been observed in the Milky Way for decades. Among them, some binaries composed of a black hole and a low-mass main sequence star -- specifically XTE J1118+480 and A0620-00 -- have exhibited anomalously fast orbital decay \cite{Gonzalez_2014}, which cannot be explained by gravitational radiation alone. The possibility that this rapid decay is caused by dynamical friction from a surrounding dark matter spike was recently explored in \cite{Chan_2023}, while the hypothesis that the black hole is of primordial origin was investigated in \cite{Ireland_2024}. While the latter study demonstrated that a PBH embedded within a DM minihalo could explain the observed decay rate, the formation mechanism of such hybrid X-ray binaries was not thoroughly addressed.

The X-ray signal is produced during the star's final few orbits around the primordial black hole, just before the merger. However, the overall merger timescale is dominated by the early stage, when the orbits are wide and long-lasting. Because of this, the merger rate given in Eq.~\eqref{eq:mergerrate} serves as a good proxy for the rate at which X-ray binary events are formed. Therefore, by estimating the merger rate within the Milky Way, we can assess whether the dynamical friction channel could be at the origin of the observed fast-decaying X-ray binaries.

To estimate this merger rate, we rely on the dynamical Milky Way model from McMillan \cite{McMillan_2017} (see also \cite{McMillan_2011} for details on the model derivation). The MW is modeled with a spherically symmetric dark matter halo, a stellar bulge, a thin and a thick stellar disk, and gas disks (see Table 3 in \cite{McMillan_2017}). Note that the DM halo in this model follows a cuspy Navarro-Frenk-White (NFW) profile.

The stellar number density $n_\ast(\mathbf{R})$ is obtained by summing the stellar mass density contributions from the bulge and the disks, divided by the mean stellar mass $\bar{m}_\ast \equiv 0.4M_\odot$. The velocity dispersion is estimated, to first order, as the circular velocity, $\sigma(R) \simeq \sqrt{GM_\text{in}(R)/R}$, where $M_\text{in}(R)$ is the total mass of the bulge and dark matter halo enclosed within radius $R$. For simplicity, we assume the bulge is spherically symmetric ($q = 1$ in Eq.~2 of \cite{McMillan_2017}) and neglect the mass contributions from the stellar and gas disks in the velocity computation. The PBH number density is obtained as $n_\text{PBH}(R)=f_\text{PBH}\times\rho_\text{DM}(R)/m_\text{PBH}$, where $\rho_\text{DM}(R)$ is provided by the DM halo profile. We fix the (subdominant) PBH fraction of DM to $f_\text{PBH}=0.01$, in broad agreement with conservative constraints on PBHs and dressed PBHs in the mass range relevant to this study \cite{Agius_2024}. 

We set the PBH mass to $6M_\odot$, approximately matching the inferred mass of the observed fast decaying X-ray binaries \cite{Gonzalez_2014}. We also fix the mass of the captured star to the mean star mass $\bar{m}_\ast$. Using this mass combination, we compute $\mathcal{I}_\text{merger}(\mathbf{R})$ (cf. Eq.~\eqref{eq:integral_mergers}) to obtain the volumetric merger rate at any location within the Milky Way. 

\begin{figure}[t]
\includegraphics[width=1.\columnwidth]{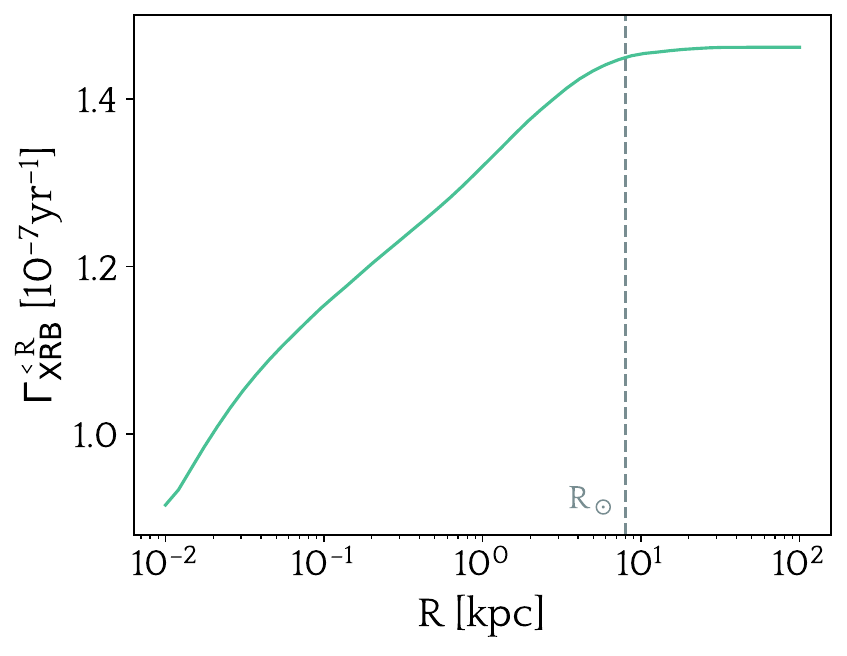}
\caption{\label{fig:formedXray} Rate of hybrid PBH-star X-ray binaries events in the Milky Way within a radius $R$ from the Galactic center, assuming $f_\text{PBH}=0.01$. The dashed line indicates the Sun's location.}
\end{figure}

This volumetric rate is then integrated over the whole volume of the Milky Way, according to Eq.~\eqref{eq:mergerrate}, to obtain the total merger rate $\Gamma_\text{XRB}=1.46 \times 10^{-7} \text{ yr}^{-1}$. The cumulative rate, within some distance $R$ from the galactic center, is shown in Fig.~\ref{fig:formedXray}. 

As seen in the figure, a significant fraction -- approximately $50\%$ -- of the total merger rate originates from the innermost region of the Milky Way, within $\sim10$ pc of the Galactic center. However, regions extending out to $\sim10$ kpc also contribute substantially to the overall rate. Thus, even if minihalos are disrupted in the central region (see Appendix \ref{app:survival}, Fig.~\ref{fig:disruption_in_MW}), a considerable number of mergers are still expected to occur at intermediate radii. Beyond $\sim10\text{ kpc}$, the stellar density becomes too low for additional systems to form. The fact that the merger rate is not entirely dominated by the central region is a consequence of the effect of perturbers (see Sec.~\ref{sec:perturbers}), which are significantly more abundant near the Galactic center and prevent most binaries from merging.


Typical low-mass X-ray binaries have X-ray emission lifetimes of the order of $10^8-10^9$ yr \cite{Paczynski_1967,Podsiadlowski_2002}, during which they may be observed. However, due to the additional orbital braking caused by dynamical friction, binaries surrounded by a DM minihalo are expected to have significantly shorter lifetimes. This can be estimated from the ratio of the orbital period to its measured time derivative, using values provided in Ref.~\cite{Gonzalez_2014}: $\tau_\text{XRB}\sim P/\dot{P}\sim\mathcal{O}(0.1\text{ day}) / \mathcal{O}(10^{-3} \text{s yr}^{-1})\sim10^7\text{ yr}$. Thus, the number of rapidly decaying X-ray binaries currently observable in the Milky Way that originate from the dynamical friction channel is expected to be $N_\text{XRB}=\Gamma_\text{XRB}\times\tau_\text{XRB}\sim\mathcal{O}(1)$. We conclude that dressed primordial black holes provide a plausible explanation for the signals observed from XTE J1118+480 and A0620-00.

It is important to note that this result is derived under the assumption that primordial black holes have a monochromatic mass distribution of $6M_\odot$ and constitute $1\%$ of the dark matter. An extended mass function for PBHs or a different abundance would lead to a different outcome. Rescaling with respect to the PBH abundance can be easily performed, given the linear dependence of the merger rate on the PBH number density.

\subsection{Gravitational wave events}
\label{sec:GWevents}

Hybrid binaries could also produce gravitational wave signals from mergers between PBHs and stellar black holes or neutron stars. In particular, they could provide an explanation to some observed events that challenge standard stellar evolution scenarios. One example is GW190521 \cite{Abbott_2020}, which involved a black hole in the upper mass gap predicted by pair-instability supernova models. Another is GW190814 \cite{Abbott_2020_bis}, characterized by a mass ratio close to $0.1$ and a compact object of $2.5M_\odot$, falling within the mass gap between neutron stars and low-mass black holes. If the PBH has a subsolar mass, it could also be the source of unusual signals such as SSM200308 \cite{Prunier_2023}.

Clearly, hybrid binaries formed through the dynamical friction channel can provide a viable explanation for these observations only if the resulting merger rate is sufficiently large. To estimate this rate, we must define again a representative volume $V$ and the corresponding $n_\ast(\mathbf{R})$, $n_\text{PBH}(\mathbf{R})$ and $\sigma(\mathbf{R})$. For simplicity, we focus on the volumetric merger rate within the local Gpc$^3$, neglecting any redshift dependence. We will estimate the merger rate in each galaxy within that volume and sum their contributions to obtain the total.

Because of the difficulty of accurately modeling the wide variety of galactic shapes and the complex distributions of dark matter and baryons across galaxies in the Universe, we adopt a simplified approach for our estimates. We assume that all galaxies are spherically symmetric. For both the DM and stellar distributions, we consider two possible profiles: a cuspy NFW and a cored Einasto profile for the dark matter, and a cuspy Jaffe and a cored Plummer profile for the stars. This yields four combinations of cored and cuspy configurations. We refer the reader to e.g.~\cite{Binney+Tremaine} for mathematical descriptions of these profiles. Given a total stellar mass $M_\ast$ for a galaxy, we associate a corresponding DM halo mass $M_\text{DM}$ using the empirical stellar mass - halo mass relation from \cite{Behroozi_2010}. We also assign a (stellar) half-mass radius, $R_{1/2}$, based on the fit for late-type galaxies from \cite{Shen_2003}, with additional support from \cite{Lange_2015} for stellar masses below $\sim 10^8M_\odot$. The halo virial radius, $R_{200}$, is then computed using the relation $R_{1/2} = 0.015 \times R_{200}$ \cite{Kravtsov_2013}. Thus, given $M_\ast$, we can determine the corresponding $R_{1/2}$, $M_\text{DM}$ and $R_{200}$, from which $n_\ast(R)$, $n_\text{PBH}(R)$, and $\sigma(R)$ can be calculated for each of the four profile combinations. For the NFW profile, we use the halo concentration from \cite{Klypin_2011}. For the Einasto profile, we fix the Einasto index to 5 \cite{Retana_Montenegro_2012} and use $M_\text{DM}$ for total halo mass and $R_{200}$ for the (DM) half-mass radius. For the Jaffe and Plummer profiles, the scale radius is determined by equating $M_\ast/2$ to the stellar mass enclosed within $R_{1/2}$, obtained by integrating the corresponding density profile. The stellar density distribution is thus a function of $M_\ast$ and $R_{1/2}$, while the dark matter distribution depends on $M_\text{DM}$ and $R_{200}$. The densities and velocity dispersions are then computed from these distributions following the same procedure as for the Milky Way in the previous subsection. In summary, all the properties of a galaxy depend only on the choice of profiles and the parameter $M_\ast$.

For a given galaxy, we compute the total merger rate $\Gamma_\text{merger}^{1\text{gal}}(M_\ast)$ (Eq.~\eqref{eq:mergerrate}) following the procedure described in Sec.~\ref{sec:mergerphasespace}. To determine the volumetric merger rate within the local Gpc$^3$, we use the galaxy stellar mass function $dn_\text{gal}/dM_\ast$ inferred from observations in \cite{Driver_2022}, which provides the number density of galaxies as a function of their stellar mass. We further integrate the total merger rate per galaxy over this distribution, considering galaxy stellar masses $M_\ast$ in the range $M_\ast^{\text{min}}=10^6M_\odot$ to $M_\ast^{\text{max}}=10^{12}M_\odot$, as follows
\begin{equation}
\label{eq:totalGWrate}
    \frac{d\Gamma_\text{merger}}{d\mathcal{V}}=\int_{M_\ast^{\text{min}}}^{M_\ast^{\text{max}}}\Gamma_\text{merger}^{1\text{gal}}(M_\ast)\frac{dn_\text{gal}}{dM_\ast}dM_\ast.
\end{equation}

We note that the total mass of galaxies within $1\text{ Gpc}^3$, computed as
\begin{align}
\begin{split}
\label{eq:totalmass}
   M_T&=\int_{M_\ast^{\text{min}}}^{M_\ast^{\text{max}}}\frac{dn_\text{gal}}{dM_\ast}(M_\ast+M_\text{DM}) dM_\ast\\
   &=5.46\times 10^{19}M_\odot 
\end{split}
\end{align}  
is in good agreement with cosmological expectations. Specifically, multiplying the (dark + baryonic) matter density parameter with the cosmological critical density and the considered volume, and further multiplying by $0.6$ (following \cite{Fukugita_2004}, where it was estimated that approximately $60\%$ of the dark matter mass is contained within the virial radii of galaxies), yields a total mass of $M_T = 2.26\times10^{19} M_\odot$, within a factor $3$ of the value from Eq.~\eqref{eq:totalmass}. This consistency ensures that the empirical laws we adopt for the stellar mass - halo mass relation and for the galaxy stellar mass function do not lead to a significant overestimation of the total galactic mass, and consequently, of the total merger rate.

To investigate the dependence of GW event rates on PBH and stellar masses, we consider several cases. We model two types of PBHs: those in the upper mass gap with a mass of $60M_\odot$, and solar-mass PBHs with $1M_\odot$. For stellar companions, we consider typical neutron stars and stellar black holes of $1.4M_\odot$ and $8M_\odot$, respectively. Since the merger rate $\Gamma_\text{merger}$ is computed using the total stellar density $n_\ast$, it must be scaled by the fraction $f$ of stars that have evolved into neutron stars or black holes to obtain the compact object merger rates (or, equivalently, the GW event rates $\Gamma_\text{GW}$), $\Gamma_\text{GW}=f\times\Gamma_\text{merger}$. We estimate this fraction as the ratio of stars between $8M_\odot$ and $25M_\odot$ for neutron stars, and above $25M_\odot$ for black holes, relative to the total number of stars, assuming a Kroupa initial mass function \cite{Kroupa_IMF}.  This yields $f\simeq0.3\%$ for neutron stars and $f\simeq0.08\%$ for black holes. Naturally, this remains a rough estimate, which assumes that all stars within these mass ranges have collapsed; a more accurate calculation would require incorporating the star formation history of each galaxy.

The total volumetric merger rates (Eq.~\eqref{eq:totalGWrate}) for the four galaxy profile combinations and the different types of mergers are presented in Table \ref{tab:merger_rates}.
\begin{table*}[t]
\renewcommand{\arraystretch}{1.5}
\begin{tabular}{|c|c|c|c|c|}
\hline
 & \; NFW + Jaffe \; & \; Einasto + Plummer \; & \; Einasto + Jaffe \; & \; NFW + Plummer \; \\ \hline
\; $1M_\odot$ PBH + $8M_\odot$ BH \; & $1.23$ & $0.47$ & $0.35$ & $0.22$ \\ \hline
\; $1M_\odot$ PBH + $1.4M_\odot$ NS \; & $0.61$ & $0.31$ & $0.21$ & $0.16$ \\ \hline
\;$60M_\odot$ PBH + $8M_\odot$ BH \; & $0.45$ & $0.25$ & $0.17$ & $0.14$ \\ \hline
\; $60M_\odot$ PBH + $1.4M_\odot$ NS \; & $0.07$ & $0.04$ & $0.03$ & $0.02$ \\ \hline
\end{tabular}
\\
\caption{\label{tab:merger_rates}Volumetric merger rates $d\Gamma_\text{GW}/d\mathcal{V}$ of PBH+BH and PBH+NS events resulting from the dynamical friction channel. All rates are expressed in units of $\text{Gpc}^{-3}\text{yr}^{-1}$. We assume $f_\text{PBH}=0.01$.}
\end{table*}

As a first observation, we see from Table~\ref{tab:merger_rates} that the merger rates are broadly comparable -- within an order of magnitude around $\mathcal{O}(0.3)\text{ Gpc}^{-3}\text{yr}^{-1}$ -- across all types of mergers. This can be understood as a consequence of competing effects: while heavier PBHs are embedded in more massive minihalos that capture stars more efficiently via dynamical friction, their number density is lower at fixed dark matter density $\rho_\text{DM}$. Similarly, stellar black holes, although more strongly affected by friction than neutron stars due to their greater mass, are intrinsically rarer. While these results were obtained assuming monochromatic PBH mass distributions, we thus expect them to remain robust for extended distributions centered around $\mathcal{O}(1-100)M_\odot$. As before, we have taken $f_\text{PBH} = 0.01$, though the rates can be rescaled straightforwardly for other PBH abundances.

\begin{figure}[t]
\includegraphics[width=1.\columnwidth]{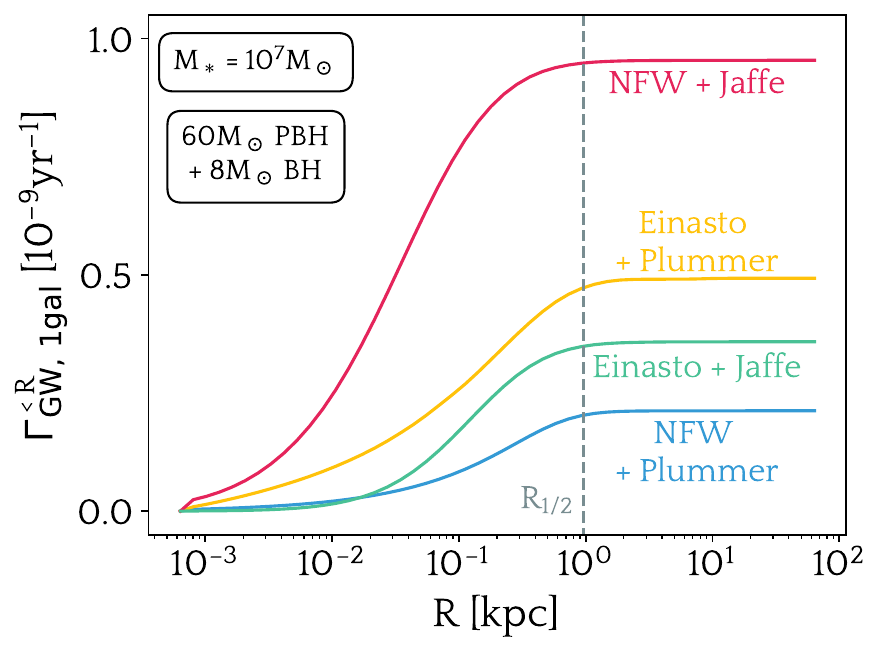}
\caption{\label{fig:rate_for_10^7Msun} Merger rate of $60M_\odot$ PBHs with $8M_\odot$ stellar BHs within a radius $R$ from the center of a galaxy with stellar mass $M_\ast = 10^7M_\odot$, assuming $f_\text{PBH} = 0.01$. Different stellar and dark matter profiles are considered. The dashed line indicates the galaxy's half-mass radius.}
\end{figure}

Another noteworthy point is that the merger rates are similar across the different combinations of galaxy density profile, highlighting the robustness of our results against variations in galaxy modeling. This behavior can be explained by the fact that, as in the Milky Way case, the merger rate in the central, extremely dense regions of galaxies with cuspy profiles is suppressed by perturbers (see Sec.\ref{sec:perturbers}), which counteracts the enhancement that would otherwise arise from the density dependence of the binary formation rate. As a result, a significant contribution to the total merger rate comes from intermediate regions of galaxies. In the absence of the perturbers condition, the rates would be dominated by the innermost regions, and cuspy profiles would lead to much higher merger rates compared to cored profiles. As an illustration, we show in Fig.~\ref{fig:rate_for_10^7Msun} the cumulative merger rate of $60M_\odot$ PBHs with $8M_\odot$ stellar BHs as a function of distance $R$ from the galactic center, for a galaxy with stellar mass $M_\ast = 10^7 M_\odot$, considering the different profile combinations.

Figure~\ref{fig:rate_for_10^7Msun} shows that the majority of the merger rate originates from the intermediate region, between approximately $0.01$ and $1$ kpc from the galactic center. Contributions from the innermost region (below $\sim0.01$ kpc) and from beyond the half-mass radius are negligible. Importantly, minihalo disruption is expected to be insignificant beyond $\sim0.01$ kpc (see Appendix~\ref{app:survival}, Fig.~\ref{fig:disruption_in_10^7Mgal}), so the merger rates are not significantly affected by DM spike disruption in galaxies.

\begin{figure}[t]
\includegraphics[width=1.\columnwidth]{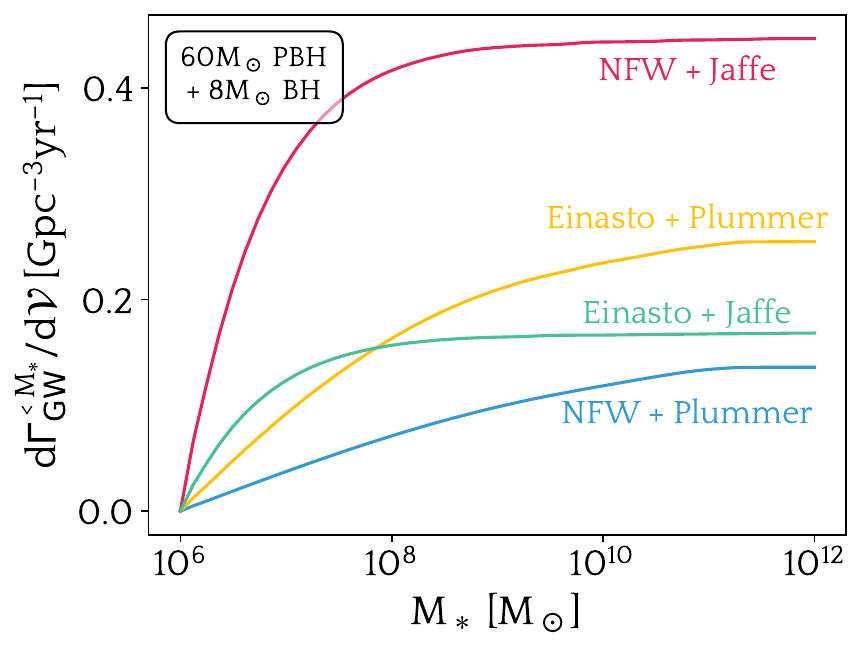}
\caption{\label{fig:rate_for_various_Mgal} Volumetric merger rate of $60M_\odot$ PBHs with $8M_\odot$ stellar BHs within galaxies with stellar mass below $M_\ast$, assuming $f_\text{PBH} = 0.01$. Different stellar and dark matter profiles are considered.}
\end{figure}

We chose to focus on a galaxy with $M_\ast = 10^7 M_\odot$ in Fig.~\ref{fig:rate_for_10^7Msun} because low-mass galaxies contribute most significantly to the overall merger rates. To further illustrate how galaxies of different masses contribute, Fig.~\ref{fig:rate_for_various_Mgal} shows the cumulative volumetric merger rate integrated over all galaxies with stellar masses below a given threshold. This is presented for the case of mergers between $60M_\odot$ PBHs and $8M_\odot$ stellar BHs, across the four galaxy profile combinations.

As shown in the figure, the galaxies that contribute most significantly to the merger rates are indeed the low-mass ones, with $M_\ast \lesssim 10^9 M_\odot$. Compared to more massive galaxies, these systems typically have denser DM halos but lower stellar densities. This turns out to be advantageous: while both stellar and dark matter densities enhance the merger rate, only the stellar component leads to disruption through the perturbers effect. As a result, although massive galaxies host the majority of the Universe's stellar mass, a distinctive feature of the dynamical friction channel is that it predominantly produces mergers in low-mass galaxies, in contrast to standard channels.

We stress, however, an important limitation of our analysis: the empirical relations used to model galactic properties -- such as the stellar mass-halo mass relation, the size-mass relation, and the galaxy stellar mass function -- are constrained by observations only for galaxies with $M_\ast \gtrsim 10^7 M_\odot$. To extend our calculation to lower masses, these relations have been extrapolated beyond their validated range. In particular, we conservatively adopted a low-mass cutoff at $M_\ast^{\text{min}} = 10^6 M_\odot$ in Eq.~\eqref{eq:totalGWrate}, given the lack of observational constraints on the galaxy stellar mass function below this scale \cite{Driver_2022}. Lowering the cutoff to, e.g., $M_\ast^{\text{min}} = 10^5 M_\odot$ increases the merger rates listed in Table \ref{tab:merger_rates} by up to an order of magnitude. Since low-mass galaxies dominate the merger rate in the dynamical friction scenario, improving both observations and modeling of faint, low-mass systems is crucial for refining these predictions. However, observations of ultrafaint dwarf galaxies within the Local Group (see e.g.~\cite{Simon_2019}) already suggest the presence of a substantial population of such galaxies. Consequently, the gravitational wave event rates estimated in this work may be significantly underestimated. 

Furthermore, primordial black holes may modify the small-scale matter power spectrum and, consequently, the halo mass function of galaxies \cite{liu2024_review,Zhang_2024}. While a detailed analysis of the interplay between PBHs and the galaxy mass function -- and its impact on the GW event rates presented here -- is beyond the scope of this work, we emphasize that our results could be significantly affected by such effects.

The rates resulting from the dynamical friction channel lie orders of magnitude below the theoretical merger rate of PBH binaries formed in the early Universe \cite{Raidal_2024}. However, the latter is highly sensitive to the initial clustering of PBHs, which remains poorly constrained, and, by definition, cannot account for observed black hole-neutron star mergers.
Previous hybrid PBH-star binary formation scenarios include direct capture via gravitational radiation and capture through three-body interactions, both of which have been found to be inefficient, yielding merger rates far below those reported here. An alternative ``catalysis'' channel, in which primordial and stellar black holes of different masses exchange binary partners to form high-mass pairs, was explored in \cite{Kritos_2021} and yielded rates comparable to the ones found here. However, this mechanism requires clustered environments and was only studied for a black hole population with three discrete masses of $10$, $60$ and $80M_\odot$.

Although the merger rates predicted in this work lie approximately one order of magnitude below the lower bound inferred from observations by the LIGO-Virgo-KAGRA collaboration \cite{LVK_2023}, they remain non-negligible. Given that some gravitational wave events are observed from sources located several Gpc away, the dynamical friction channel could still account for a few tens of detectable events per year. It therefore offers a plausible explanation for the exotic GW events discussed at the beginning of this subsection.

\section{Conclusion}
\label{sec:conclusion}

In this work, we investigated a novel channel for the formation of ``astro-primordial'' hybrid binary systems, consisting of a primordial black hole and a stellar object of any kind. This formation mechanism relies on the presence of a dark matter spike surrounding the black hole. As stars pass through the minihalo, they experience dynamical friction from the DM particles, which slows them down and can lead to gravitational capture. We computed the binary formation rate both analytically and numerically by solving the equations of motion for a test particle inside the minihalo. By further restricting to binaries that lose sufficient energy during the subsequent crossings to merge within a finite time -- despite the presence of external perturbers such as passing stars -- we derived the corresponding merger rates. We applied these results to two observational contexts: X-ray binaries in the Milky Way with anomalously rapid orbital decay, and compact object mergers responsible for gravitational wave events. We found that this mechanism offers a plausible explanation for several puzzling observations from recent years. Finally, we showed in the Appendix that minihalo disruption over their lifetimes primarily affects those in the innermost regions of galaxies, and thus does not significantly affect our conclusions. Clearly, the merger rates derived in this work are order-of-magnitude estimates, with ample room for improvement falling into two main categories: galaxy modeling and dynamical friction modeling.

Regarding galaxy modeling, our analysis relied on simplified, spherically symmetric galaxy models (with the exception of the Milky Way), adopting standard profiles for the stellar and dark matter distributions. We found that mergers induced by dynamical friction predominantly originate from the intermediate regions of low-mass galaxies -- a notable difference with standard scenarios, which typically associate mergers with massive galaxies or dense, clustered environments. However, our modeling assumed specific, fixed relations between stellar mass, dark matter mass, half-mass radius, and virial radius. In reality, galaxies exhibit a wide distribution of these properties. A more refined treatment would require incorporating the full diversity of galaxy populations, with particular focus on the small, dark matter-dominated dwarf galaxies that remain poorly constrained by observations, and whose properties were extrapolated from those of more massive systems. Additionally, our analysis did not account for the evolution of galaxies or their stellar populations with redshift. Incorporating realistic star formation histories would be essential to obtain more accurate and time-dependent merger rate predictions. The stellar evolution into compact objects -- and particularly the natal kicks -- could also influence the outcomes. Similarly, accounting for a nonmonochromatic population of PBHs and for the low-redshift growth and/or partial disruption of their surrounding minihalos -- depending on their location within galaxies -- would further enhance the accuracy of the results.

Concerning dynamical friction, a key assumption in our analysis is that the minihalos remain static over the lifetime of the binary. In reality, while the stellar mass is small relative to the spike, it can reach up to $\sim10\%$ of the dark matter mass depending on the merger scenario. As a result, the minihalo may be dynamically heated as the star spirals inward. A comparison between the minihalo binding energy (Eq.~\eqref{eq:bindingE}) and the typical orbital energy of the binary system $\sim Gm_\ast m_\text{PBH}/r$ indicates that the energy injected into the spike by dynamical friction becomes comparable to the minihalo binding energy once the separation $r$ between the two bodies falls below $\mathcal{O}(10)$ AU. Beyond this point, the further evolution of the system is uncertain, and a precise analysis -- accounting for the evolution of the spike and energy loss via gravitational radiation (the latter will become relevant at $r\lesssim 0.01$ AU, see e.g. Eq.~(5.10) of \cite{Peters_1964}) -- is required. We note that the full co-evolution of the binary and the surrounding minihalo has been studied in detail -- in the large mass-ratio limit of the merging objects -- in Ref.~\cite{Kavanagh_2020}, and we leave a corresponding analysis in the present context to future work. On the other hand, the presence of a DM spike around compact object binaries can lead to observable dephasing in gravitational waveforms, potentially serving as a smoking gun for identifying such minihalos \cite{Eda_2015,Coogan_2022}. Finally, we have assumed a noninteracting dark matter candidate in the spike; for models where self-annihilation is relevant, the central density may be capped, potentially modifying the spike structure and thus the effectiveness of the dynamical friction channel.

Among the potential future directions, it would be interesting to assess whether the dynamical friction channel could enhance the merger rate of PBH binaries. While the survival of DM minihalos around early-forming PBH binaries in clustered environments requires careful treatment, a potentially significant fraction of PBH binaries might instead form at late times in nonclustered regions, where minihalos are more likely to survive and the dynamical friction channel may apply. Another particularly intriguing prospect is the study of dressed primordial black holes interacting with binary star systems. Such interactions could lead to nonstandard pathways of binary stellar evolution and may represent a significant channel for second-generation mergers. 

\acknowledgements{
It is a pleasure to thank Dominic Agius, Irina Dvorkin, Ga\'etan Facchinetti, Nicolas Grimbaum Yamamoto, Aurora Ireland, Nour Mouhssine, and Marine Prunier for interesting discussions. Peter Tinyakov deserves special thanks for his comments on the manuscript and the many valuable discussions. We also thank the anonymous referee for their valuable comments, which helped improve the quality of this work. The author is a FRIA grantee of the Fonds de la Recherche Scientifique-FNRS. Tag: ULB-TH/25-07}

\bibliographystyle{apsrev4-2} 

\bibliography{bibli}

\appendix
\section{Survival of the minihalos}
\label{app:survival}

An important effect that must be considered is the possible disruption of the DM spikes between their time of formation and the capture of a star from the environment at recent times. In this appendix, we provide quantitative arguments showing that the effect of minihalo disruption on binary formation and merger rates is mild and can be safely neglected. Following Ref.~\cite{Hertzberg_2020}, we examine three mechanisms that could potentially disrupt minihalos: global tides from the galactic halo, high-speed stellar encounters, and disk shocking. 

\subsection{Global tides}
An object of finite size, such as a DM spike, will be subject to the static tidal field of the galactic halo in which it is embedded. In the distant-tide approximation and assuming a spherically symmetric galaxy, one finds the associated tidal radius \cite{Binney+Tremaine,Hertzberg_2020}
\begin{equation}
\label{eq:globaltide}
    r_t=R\left(\frac{m_\text{dPBH}}{3M_\text{gal}(R)}\right)^{1/3}\left(1-\frac{1}{3}\left.\frac{d\ln M_\text{gal}(\mathcal{R})}{d\ln \mathcal{R}}\right|_{R}\right)^{-1/3},
\end{equation}
where $M_\text{gal}(R)$ denotes the total mass enclosed within galactocentric radius $R$. The tidal radius sets an upper bound on the minihalo size: particles at distances beyond $r_t$ experience tidal forces from the galaxy that exceed the minihalo self-gravity. Both the tidal radius and the minihalo radius scale as $m_\text{PBH}^{1/3}$ (see Eqs.~\eqref{eq:radspike} and ~\eqref{eq:globaltide}). As a result, the critical galactocentric distance $R_c$ below which $r_t<r_\text{sp}$ is independent of the PBH mass. 

Using the Milky Way model described in Sec.~\ref{sec:Xray}, and considering only the contributions from the stellar bulge and DM halo, we find $R_c=0.48\text{ kpc}$. For the various galactic models discussed in Sec.~\ref{sec:GWevents}, the corresponding value of $R_c$ ranges from $0$ to $0.029\text{ kpc}$ for a galaxy with stellar mass $10^7M_\odot$, with larger $R_c$ values found in models featuring cuspy (Jaffe) stellar profiles.

\subsection{Minihalo binding energy}
The stability of the minihalo is maintained by its self-gravitational binding energy. Suppose that some of its outer layers have already been removed -- due to, for instance, the global tides discussed in the previous subsection -- leaving it with a current radius $r_i \leq r_\text{sp}$. The energy required to further unbind all mass shells between $r_i$ and a smaller radius $r_f < r_i$ is given by:

\begin{equation}
        E_b=4\pi G\int^{r_i}_{r_f}r\rho_\text{sp}(r)\left[m_\text{sp}(r)+m_\text{PBH}\right]dr.
\end{equation}
For $r_f\gg r_s$, where $r_s=2Gm_\text{PBH}/c^2$ is the Schwarzschild radius of the central PBH, the dominant contribution arises from the mass of the spike itself. Using Eq.~\eqref{eq:radspike} and \eqref{eq:rhospike}, one then finds

\begin{align}
\begin{split}
    \label{eq:bindingE}
    E_b&\simeq1.5\times10^{-10}\left(\frac{m_\text{PBH}}{M_\odot}\right)^{5/3}M_\odot c^2\\
    &\times\left(\frac{r_i}{r_\text{sp}}\right)^{1/2}\left[1-\left(\frac{r_f}{r_i}\right)^{1/2}\right].
\end{split}
\end{align}

By comparing this energy with the energy injected into the minihalo through high-speed stellar encounters and disk shocking, we can estimate the impact of these processes on the minihalo size. We now examine each mechanism in detail.

\subsection{High-speed encounters with stars}
\label{sec:highspeedstars}

While star-minihalo interactions leading to binary formation typically occur at low relative velocities, the majority of stellar encounters will involve high-velocity stars. These fast-moving stars, as they pass through or near the minihalo, will inject energy into it, potentially altering its internal structure. In the impulse approximation, the energy injected by a single star of mass $m_\ast$ passing through or near a minihalo with impact parameter $b$ and velocity $\sigma_\text{rel}$ is \cite{Carr_1999,Hertzberg_2020}

\begin{align}
\begin{split}
\label{eq:onestar_disruption}
    E_\ast^1&\simeq\frac{16\pi}{3}\left(\frac{G m_\ast}{\sigma_\text{rel}b^2}\right)^2\\ &\times 
    \int_{4r_s}^{r_i}r^4\rho_\text{sp}(r)\left(1+\frac{4r^4}{9b^4}\right)\left(1+\frac{2r^2}{3b^2}\right)^{-4}dr,
\end{split}
\end{align}
where we integrate over the whole minihalo, with the lower limit $4r_s$ being set by the radius below which particles from the spike are captured by the black hole \cite{Gondolo_1999}. For the spike profile given in Eq.~\eqref{eq:rhospike}, the integral can be evaluated analytically and expressed in terms of trigonometric functions. 

The critical impact parameter $b_c$, below which a minihalo is fully disrupted by a single stellar encounter (a so-called ``one-off'' disruption), can be computed by equating Eq.~\eqref{eq:onestar_disruption} with Eq.~\eqref{eq:bindingE}, evaluated at $r_i = r_\text{sp}$ and $r_f = 0$. For typical values of $m_\text{PBH}=1M_\odot$, $m_\ast=0.4M_\odot$ and $\sigma_\text{rel}=200\text{ km/s}$, one finds $b_c=1.6\times10^{-8} \text{ pc}$. The corresponding number of encounters with $b<b_c$ is then given by
\begin{align}
\begin{split}
    N_\text{one-off}&=n_\ast \sigma_\text{rel} T_u\pi  b_c^2\\
    &=1.7\times 10^{-9} \left(\frac{n_\ast}{\text{pc}^{-3}}\right)\left(\frac{T_u}{10^{10}\text{yr}}\right).
\end{split}
\end{align}
Therefore, the probability of one-off disruption is entirely negligible in any realistic galactic environment.

However, cumulative energy injection from numerous encounters with $b > b_c$ can still significantly affect the minihalo structure. The total energy deposited into the DM spike can be calculated by summing the contributions from all such encounters:
\begin{equation}
\label{eq:totalstarinput}
    E_{\ast}^\text{tot}=2\pi\sigma_\text{rel}T_un_\ast\int_{b_c}^\infty bE_\ast^1db.
\end{equation}
The evolution of the minihalo radius is obtained by equating Eq.~\eqref{eq:totalstarinput} with Eq.~\eqref{eq:bindingE} and solving for $r_f$.

\subsection{Disk shocking}
\label{sec:diskshock}
Rather than adopting a localized approach, where individual stars inject energy into the minihalo, one can take a complementary global approach by estimating the total energy injected into the minihalo as it crosses the galactic stellar disk. The energy per unit mass injected into DM particles located at a distance $r$ from the center of the minihalo during a single disk crossing is given by \cite{Hertzberg_2020,Stref_2017,Binney+Tremaine}
\begin{equation}
\label{eq:E_1DS}
    \mathcal{E}_\text{DS}^1=\frac{32\pi^2G^2\rho_d^2z_d^2r^2}{3v_z^2}A(\eta),
\end{equation}
where $\rho_d$ and $z_d$ denote the local mass density and vertical scale height of the disk. $v_z$ is the vertical velocity of the minihalo relative to the (horizontal) disk, assumed to be constant during the crossing and related to the local circular velocity around the galactic center as $v_z=\sigma/\sqrt{2}$. $A(\eta)$ is an adiabatic correction factor that accounts for deviations from the impulse approximation. In particular, $A(\eta)$ suppresses energy injection into the minihalo when the orbital period of its particles around the central PBH becomes shorter than the disk crossing time. It is defined as $A(\eta)=(1+\eta^2)^{-3/2}$, with $\eta=\omega\times\tau$ the adiabatic parameter, $\omega=[G(m_\text{sp}(r)+m_\text{PBH})/r^3]^{1/2}$ the circular velocity of DM particles around the central PBH and $\tau=H/v_z$ the disk crossing time. $H$ is the effective half-height of the disk, related to the scale height $z_d$, in the case of an exponentially decreasing profile, as $H=\ln(2)\times z_d$.

The energy injected into the minihalo in one disk crossing is obtained by integrating Eq.~\eqref{eq:E_1DS} over the mass elements of the minihalo
\begin{equation}
\label{eq:ET_1DS}
E_\text{DS}^1(r_i)=4\pi\int_{4r_s}^{r_i}r^2\rho_\text{sp}(r)\mathcal{E}^1_\text{DS}(r)dr.
\end{equation}
On the other hand, the number of disk crossings, in a time $T_u=10^{10}$ yr, can be estimated as the ratio between this time and half the orbital period, $N_c=T_uv_z/(\pi R)$.

Importantly, after each crossing, the injected energy results in a removal of the outer layers of the minihalo, thereby reducing its radius from $r_i$ to $r_f$. However, as seen from Eq.~\eqref{eq:ET_1DS}, the energy injected per crossing decreases as the minihalo shrinks. Consequently, the total energy injected over $N_c$ crossings must be computed iteratively: starting from an initial radius $r_i$, one computes the injected energy via Eq.~\eqref{eq:ET_1DS}, then updates the radius using Eq.~\eqref{eq:bindingE} to obtain $r_f$. This new radius is used as the input $r_i$ for the next crossing, and the process is repeated until all $N_c$ crossings are accounted for.

\subsection{Combined effect \& discussion}

We now look at the combined effect of global tides and stellar interactions on minihalos. We consider the Milky Way model described in Sec.~\ref{sec:Xray}, as well as a $M_\ast=10^7M_\odot$ galaxy with the stellar and DM profiles introduced in Sec.~\ref{sec:GWevents}. At each galactocentric distance $R$, we compute the tidal radius $r_t$ (Eq.~\eqref{eq:globaltide}) and set the initial minihalo size to $\min(r_t,r_\text{sp})$. This radius is then further evolved using the procedure outlined in Sec.~\ref{sec:highspeedstars} for high-speed stellar encounters or Sec.~\ref{sec:diskshock} for disk shocking. The latter mechanism is only applied to the Milky Way model, as the other galaxies are modeled without disks. For the Milky Way, we conservatively adopt stellar densities in the galactic plane for the calculation of encounter rates. Stellar and PBH masses are fixed to $0.4M_\odot$ and $1M_\odot$, respectively, and we verified that our conclusions hold across the full PBH mass range considered in this work.

\subsubsection{Milky Way}

\begin{figure}[t]
\includegraphics[width=1.\columnwidth]{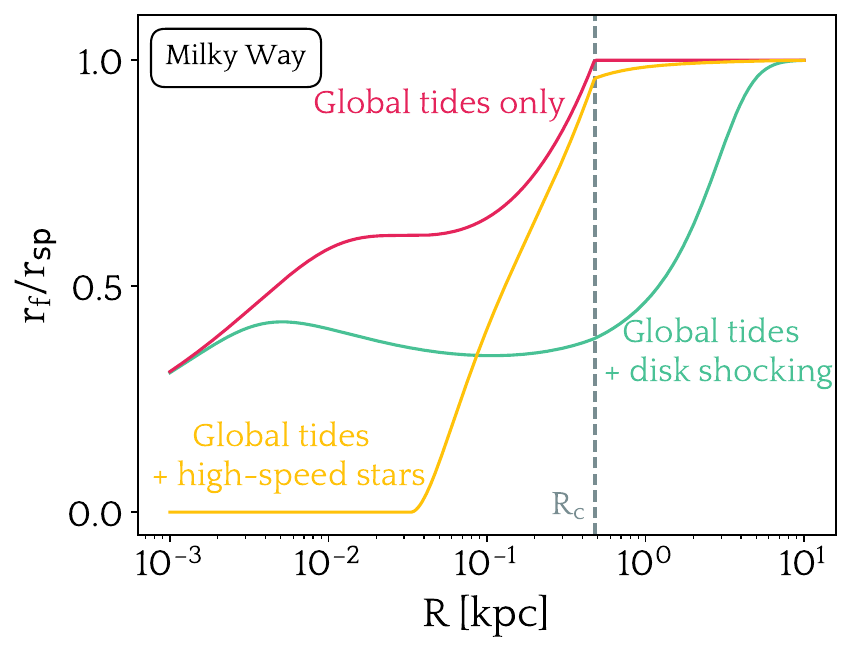}
\caption{\label{fig:disruption_in_MW} Ratio of the final to initial minihalo radius after disruption via various mechanisms, as a function of the distance $R$ from the center of the Milky Way. The dashed line indicates $R_c$, the critical radius below which $r_t<r_\text{sp}$.}
\end{figure}

In Fig.~\ref{fig:disruption_in_MW}, we focus on the Milky Way and plot the ratio of the final minihalo radius to its original, unstripped value given by Eq.~\eqref{eq:radspike}, as a function of the minihalo's galactocentric distance $R$.

In this figure, the upper red curve shows the minihalo radius when only global tides from the galactic halo are considered. A sharp transition appears at $R_c=0.48\text{ kpc}$, where $r_t=r_\text{sp}$. The central yellow curve includes the effect of individual high-speed stellar encounters, while the lower green curve accounts for energy injection from disk shocking.

Disk shocking does not fully disrupt the minihalos. This is because we account for the gradual evolution of the minihalo radius, which suppresses further disruption: as the radius shrinks, the injected energy per crossing decreases, limiting further stripping. The upturn of the red curve below $\sim0.01\text{ kpc}$ reflects the adiabatic correction to the impulse approximation. As $v_z$ decreases closer to the center, disk crossings become slower, making the interaction more adiabatic and reducing energy injection.

On the other hand, the trend of the yellow curve is straightforward: as the galactocentric distance $R$ decreases, the increasing stellar density enhances energy injection by high-speed stars into the minihalo, eventually exceeding its total binding energy. As a result, minihalos are completely disrupted below $\sim 0.03\text{ kpc}$.

Although a more refined treatment of stellar encounters would involve tracking the minihalo radius after each encounter -- analogous to the approach used for disk shocking -- the current conservative estimates are sufficient for our purposes. Above $\sim 0.03\text{ kpc}$, minihalos are only partially disrupted, with their radii reduced by at most an $\mathcal{O}(1)$ factor. According to Eq.~\eqref{eq:analytical_phasespace}, the binary formation rate scales with the minihalo mass $m_\text{sp}$, which for the DM density profile of Eq.~\eqref{eq:rhospike}, behaves as $m_\text{sp}\propto r^{3/4}$. Moreover, binaries that ultimately merge are those formed from stars passing close to the center of the DM spike, where dynamical friction is most effective. As a result, the merger rate is expected to depend even more weakly on the outer radius of the minihalo. Consequently, an $\mathcal{O}(1)$ reduction in minihalo radius leads to at most an $\mathcal{O}(1)$ suppression on the merger rates. Since a substancial fraction of X-ray binaries are formed at radii $R\gtrsim 0.03\text{ kpc}$ (see Fig.~\ref{fig:formedXray}), we conclude that DM spike disruption does not significantly affect the overall X-ray binary formation rate.

We note that the iterative evolution of minihalos assumes sufficient time for revirialization between subsequent disk crossings. A comparison between the free-fall timescale of DM particles in the spikes and half the orbital period of their central PBHs around the galactic center indicates that this condition is satisfied beyond $\sim0.01$ kpc, supporting the validity of the iterative procedure in that region.

\subsubsection{$M_\ast=10^7M_\odot$ spherical galaxy}
\begin{figure}[t]
\includegraphics[width=1.\columnwidth]{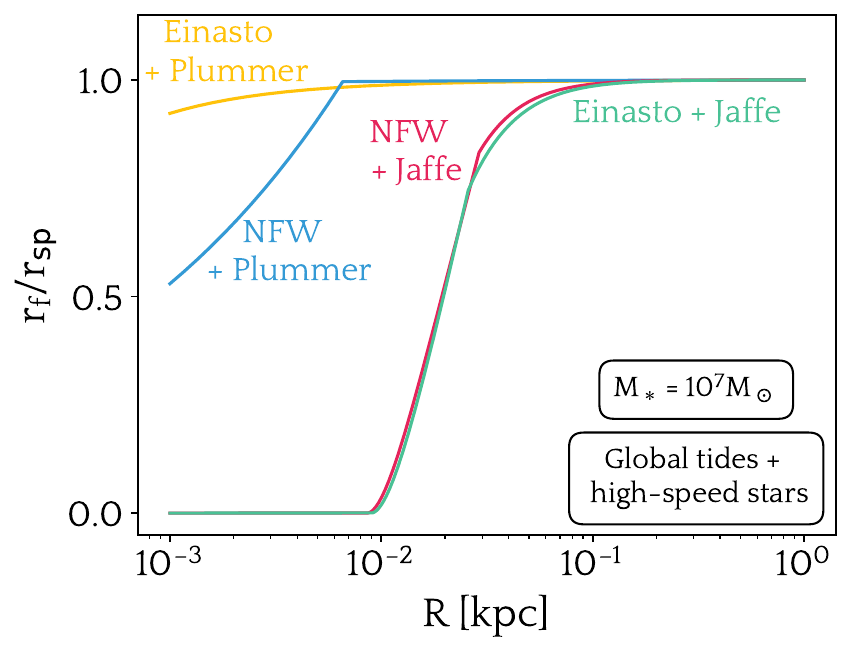}
\caption{\label{fig:disruption_in_10^7Mgal} Ratio of the final to initial minihalo radius after disruption via global tides and fast stellar encounters, as a function of the distance $R$ from the center of a $M_\ast=10^7M_\odot$ galaxy, assuming various stellar and DM density profiles.}
\end{figure}

We now turn to the spherical galaxy models used to estimate the GW event rate. For these systems, we consider only global tides and individual stellar encounters. The combined effects of these two processes are shown in Fig.~\ref{fig:disruption_in_10^7Mgal} for the various combinations of stellar and dark matter density profiles.

Minihalos in galaxies with cored (Plummer) stellar profiles are essentially unaffected by stellar encounters, as the stellar densities remain too low even in their central regions. Global tides, however, have a mild impact when the dark matter profile is cuspy (NFW), as illustrated by the blue curve. In contrast, galaxies with cuspy (Jaffe) stellar profiles exhibit significant minihalo disruption in their central regions, with DM spikes not surviving below $\sim 0.01\text{ kpc}$ from the galactic center. Nevertheless, as shown in Fig.~\ref{fig:rate_for_10^7Msun}, most dynamical friction-induced mergers occur at radii larger than this. We therefore conclude that minihalo disruption does not significantly affect merger rates in typical $M_\ast=10^7M_\odot$ galaxies. Moreover, since galaxies in this mass range dominate the contribution to the gravitational wave signal (see Fig.~\ref{fig:rate_for_various_Mgal}), we conclude that the overall GW event rate is not substantially impacted by minihalo disruption.

As a final remark, we note that throughout this Appendix, we have assumed that minihalos follow circular orbits at fixed galactocentric distance $R$. A more realistic treatment would require accounting for eccentric orbits, which dressed PBHs are likely to follow around galactic centers.

\end{document}